\documentclass[letterpaper]{article}

\usepackage[T1]{fontenc}

\usepackage{geometry}
\usepackage{setspace}
\usepackage{gensymb}
\usepackage{textcomp} 
\usepackage{achemso}

\usepackage{graphicx}
\usepackage{float}
\newfloat{scheme}{htbp}{los}
\floatname{scheme}{Scheme}
\floatname{chart}{Chart}
\newfloat{graph}{htbp}{loh}

\usepackage{chemformula} 
\usepackage[version = 4]{mhchem} 
\usepackage{array}
\usepackage{booktabs}
\usepackage{multirow}
\usepackage{caption}

\usepackage{authblk}
\author[1]{Fernando M. Fernandes*}
\affil[1]{IMCN/NAPS, Universit\'e catholique de Louvain, Louvain-la-Neuve, Belgium}
\author[2]{Fouad El Haj Hassan}
\affil[2]{Basic and Applied Sciences Research Center, Al Maaref University, Beirut, Lebanon}
\author[3]{Sophie Hermans}
\affil[3]{IMCN/MOST, Universit\'e catholique de Louvain, Louvain-la-Neuve, Belgium}
\author[1]{Beno\^it Hackens}

\title{Elucidating different \ch{NO2} sensing mechanisms in oxidized PbS nanocrystals}

\date{*Email: fernando.massa@uclouvain.be}

\begin{document}

\maketitle

\begin{abstract}


In this work we provide an in-depth analysis of the sensing mechanisms of \ch{NO2} by lead-sulfide nanocrystals (PbS-NCs). 
A detailed model for the sorption mechanism is proposed, and the correlation is established between experimental sensing characteristics and the surface composition, based on both experimental characterization and ab initio (DFT) simulations.
We demonstrated how the sensitivity and the sensing dynamic response can be tuned by a post-deposition multistep dry-thermal process at mild temperature, that alternates vacuum-assisted annealing and heating in open-air. Sensors with different surface compositions were fabricated, and their dynamic response was characterized at low concentration of \ch{NO2} (0.5 ppm) in air, at ambient temperature. DFT simulations indicate that both surface stoichiometry and oxidation critically govern \ch{NO2} interaction on \ch{PbS}, with sulfur-rich terminations favoring weaker binding and faster desorption, while intermediate oxidation enhances interaction and over-oxidation leads to surface passivation, in agreement with the measured experimental sensing dynamics. By linking surface composition, adsorption chemistry, and resistance transduction within a single framework, this work provides clear indications to design room-temperature, low-ppm \ch{NO2} microsensors fabricated through a simple and scalable processes.

\end{abstract}

\section*{Keywords}

Gas sensor, sensing mechanism, nanoparticles, DFT, environmental monitoring.




\section{Introduction}


Gas microsensors play a central role in a diverse set of smart applications including real-time environmental monitoring, indoor air-quality control, healthcare, agricultural/food quality control, public safety (e.g. potential explosions, chemical hazards, fire warning, etc) and leak detection in industrial settings \cite{Zong2025,Mandal2025,Wang2025,Acharyya2024,Upadhyay2024}. In particular, planar thin-film gas sensors, based on conductometric principle (i.e. chemiresistive type), offer great potential for miniaturisation and mass-production by wafer-level planar integration and batch processing using modern semiconductor manufacturing technologies (CMOS, complementary metal-oxide-semiconductor) \cite{Xu2015,Song2021,Filipovic2022,Lakshminarayana2025,Luo2025}. This type of sensors can be simply manufactured by depositing the sensitive layer on top of a pair of interdigitated electrodes (IDEs) that can be pre-patterned on an insulating substrate (typically, alumina or silicon oxide)\cite{Korotcenkov2013}. Traditionally, the sensing layer is based on semiconductor metal oxides (SMOs), including \ch{ZnO}, \ch{SnO2}, \ch{In2O3}, \ch{MoO3}, \ch{Mn3O4}, and \ch{WO3}, and aim to detect reducing or oxidizing gases like \ch{CO}, \ch{NO2}, \ch{H2S}, \ch{H2}, \ch{NH3}, \ch{CH4}, or volatile organic compounds (VOCs) \cite{Neri2015,PriyaGupta2025}. However, a common issue in sensing reactive gases, in particular strong oxidants like \ch{NO2}, is the requirement of high operational temperature (200-250$^{\circ}$C) in order to improve performance and accelerate gas desorption (recovery) after exposure. \ch{NO2} is one of the most dangerous air pollutants that can impact human health even at very small concentrations (e.g., the exposure limit in working place, for an 8h-shift, was set to 0.5 ppm, part-per-million, by the EU Commission Directive 2017/164/EU). Therefore, in recent years, great effort has been dedicated to the design of new materials to enhance the overall sensor efficiency when operated at ambient temperature. The main strategies generally involve chemical sensitization with noble metal traces (e.g. Pt, Au, Pd), the application of nanomaterials (e.g. nanoparticles, nanorods, mesoporous sheet), and hybridization with conductive polymers and 2D-materials \cite{Bulemo2025,Filipovic2022,Saruhan2023,Yun2022,Zhou2021,Xuan2020,WANG2023}.

When exposed to the atmosphere, gas and water molecules adsorb on the sensor surface and diffuse through its porosity, eventually saturating the sensing layer. While gas diffusion depends on synthesis and deposition conditions, sensing in SMOs is most commonly governed by oxygen adsorption: oxygen molecules occupy vacancies, forming negatively charged species (\ch{O^-_{2}}, \ch{O^-}, or \ch{O^2-}) that trap electrons in n-type SMOs (e.g., \ch{SnO2}) or donate holes in p-type SMOs (e.g., \ch{Mn3O4}). Humidity effects mainly arise from dissociative \ch{H2O} adsorption, where hydroxyl groups interact with lattice metals and generally act as electron donors in n-type SMOs. These surface processes tune the bulk Fermi level, leading to resistance changes that increase in n-type or decrease in p-type materials. In the presence of reducing (e.g., \ch{NH3}) or oxidizing (e.g., \ch{NO2}) gases, reactions with adsorbed oxygen release or replace these species, respectively. In nanocrystalline grains with sizes comparable to the Debye length, modulation of electron depletion or hole accumulation layers strongly enhances sensitivity, as does tuning of interfacial potential barriers such as double Schottky barriers. Nevertheless, a clear correlation between sensing response and sensing-layer composition remains elusive due to the difficulty of precisely defining surface reactions and sorption dynamics.

As a simple alternative to SMOs, the gas-sensing capabilities of lead-sulphide nanocrystals (PbS-NCs) at ambient temperature, has been investigated towards reducing or oxidizing gases depending on the specific surface chemistry \cite{Mitri2020,Li2022,Kwon2024,Liu2015,Liu2016,Mosahebfard2016}. In simple PbS-NCs, the conducting cores are encapsulated by an oxidized surface layer that is reactive to \ch{NO2}. The sensing layer is formed by  an ensemble of connected NCs in which conductivity is limited by the potential barriers formed at their interfaces. In this work we perform a detailed analysis of the mechanisms involved in \ch{NO2} gas sensing, based on a sensing layer formed of PbS-NCs. The sensing layers in our devices were formed after deposition by combining different steps of vacuum-assisted annealing and heat-treatment in open air, for layer crystallization and surface modification. The crystallinity and surface composition after the thermal treatment steps were characterized, and the reactivity to \ch{NO2} was analyzed with the support of DFT calculations; using model-particles featuring different elemental amounts of lead, sulfur and oxygen atoms. Those results allowed us to propose a detailed surface reaction model and establish the correlation between the experimental dynamic response, the lead-to-sulfur ratio, and the ratio amounts of lead in different oxidation states (\ch{Pb^0}, \ch{Pb^{2+}}, \ch{Pb^{4+}}).

\section{Results and discussion}

\subsubsection*{Sensors fabrication}\label{Fabrication}

\begin{figure}[h!]
\centering
\begin{tabular*}{\textwidth}{ll}
{$\mathrm{\textbf{a \ }}$}{\includegraphics[height=0.275\columnwidth]{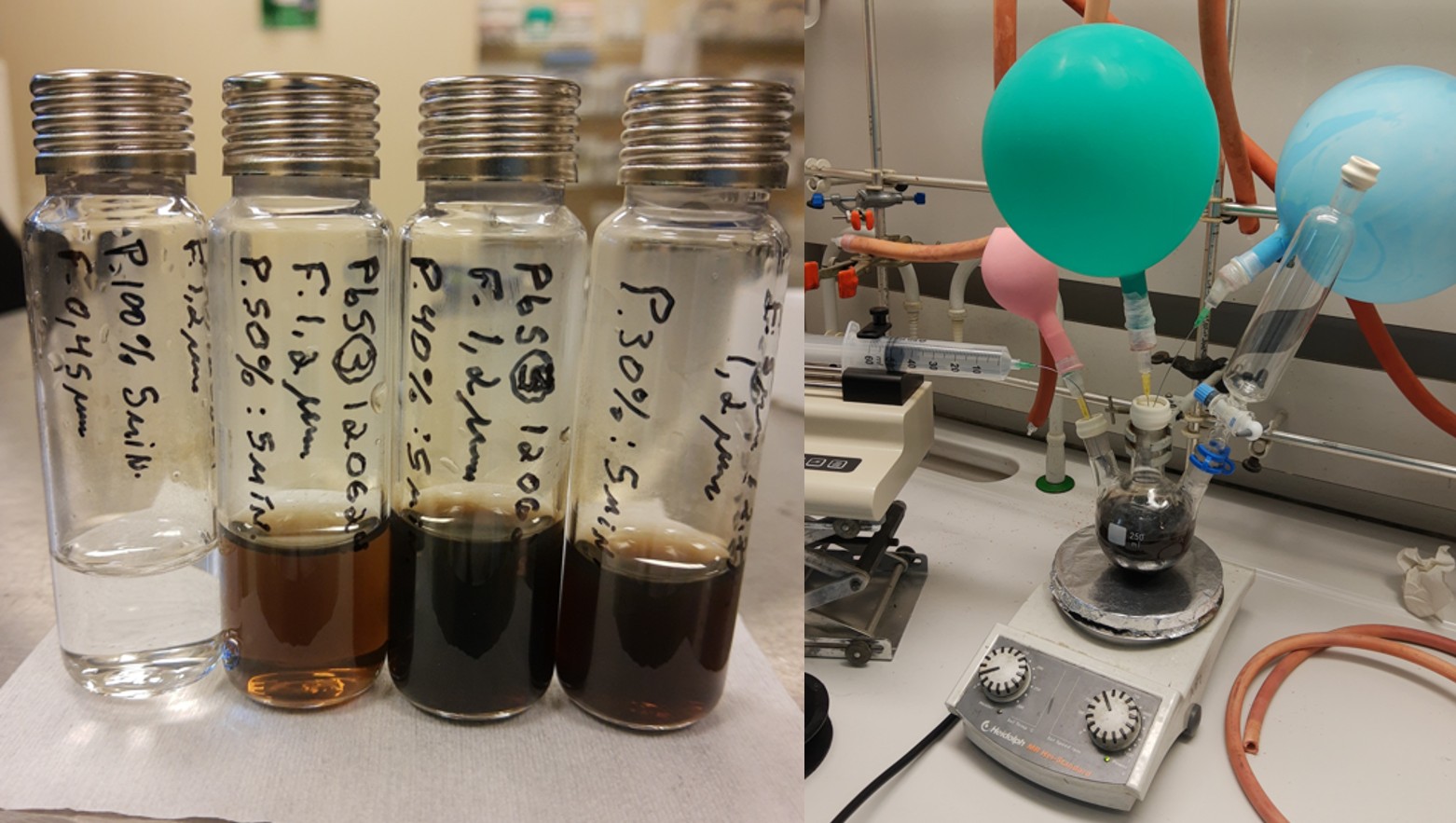}}
 & {$\mathrm{\textbf{b \ }}$}{\includegraphics[height=0.275\columnwidth]{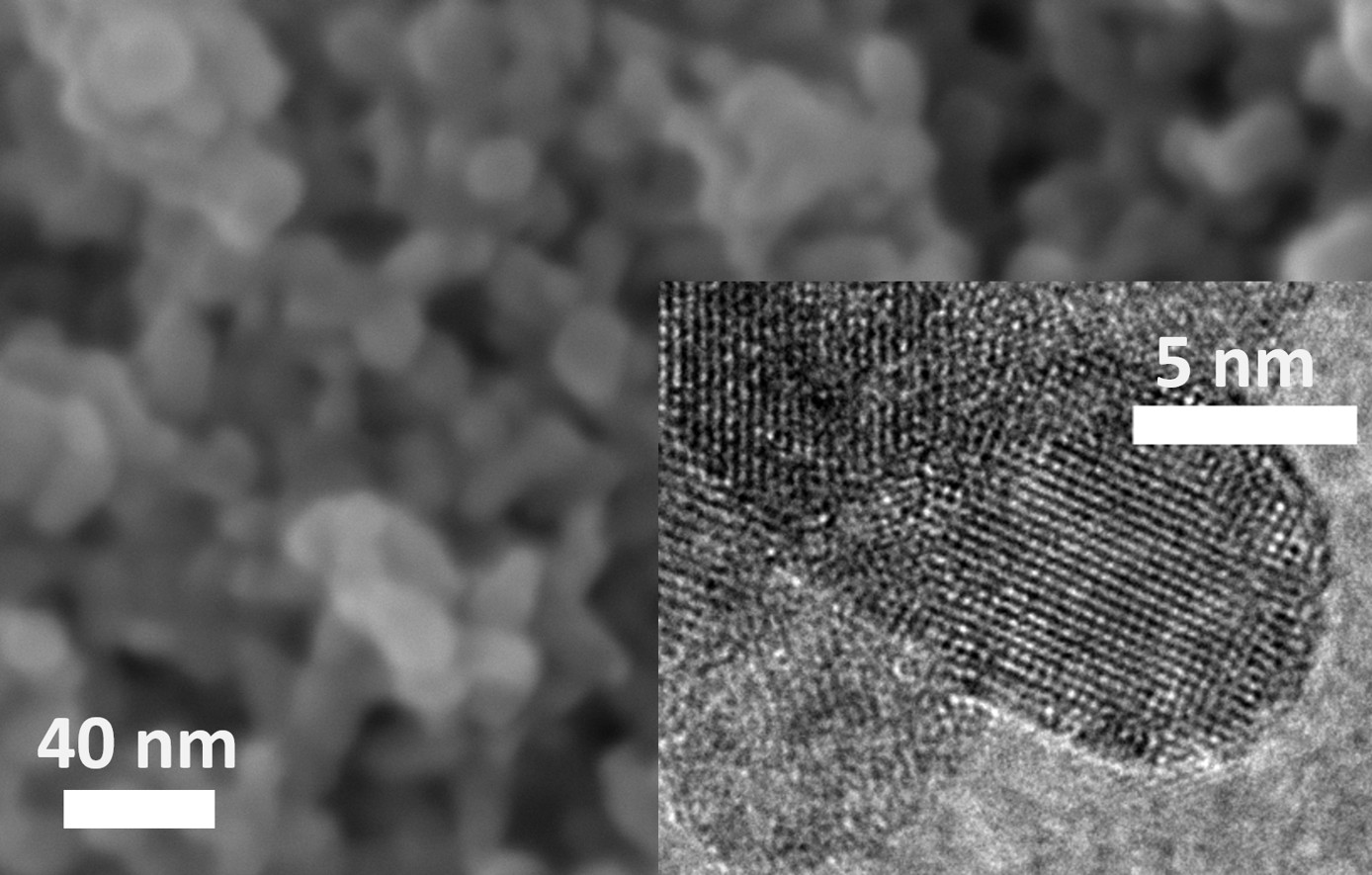}} \\ \\
 \end{tabular*}
 \begin{tabular*}{\textwidth}{c}
{$\mathrm{\textbf{c \ }}$}{\includegraphics[height=0.30\columnwidth]{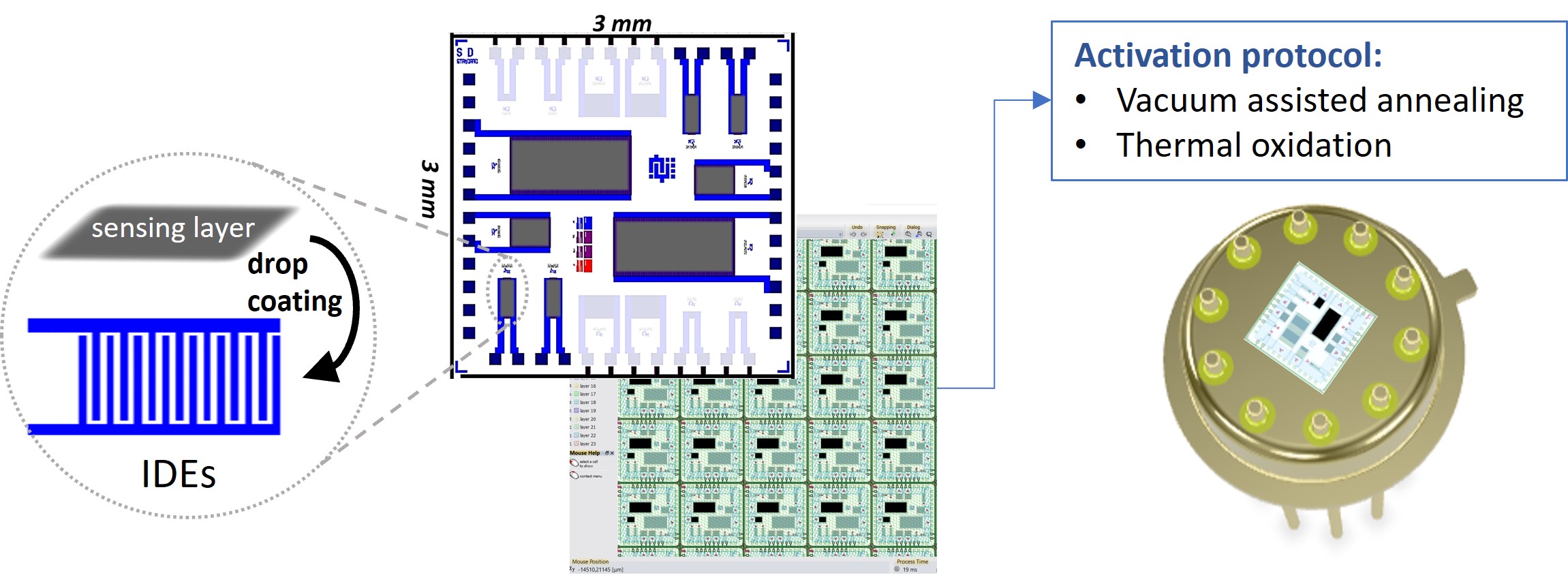}} \\
\end{tabular*}
\caption{Fabrication: (a) \textit{On the left}, the PbS-NPs dispersed in water (ink) obtained after ultra-sonication at 4-different powers (from left-to-right: 100 \%, 50 \%, 40 \% and 30 \%) and filtering to discard largest agglomerates. \textit{On the right, in panel (a)}, the simple experimental setup used for the synthesis of PbS-NPs, at room temperature and using water as the main solvent \cite{Nabiyouni2012}. (b) SEM-image showing the morphology of the sensing layer of PbS-NPs after the heating step in vacuum at 180 $\celsius$ for 30 minutes. In the \textit{inset}, the high-resolution (TEM) image showing a single PbS nanoparticles with its crystalline core. (c) The scheme of the fabrication process, with the deposition of the sensing layer by drop coating, the geometry of the IDEs-array chip, the wafer-printing layout, and the illustration of the final device.}\label{Fig.Fabr}
\end{figure}

A simple fabrication method was applied based on the drop casting technique, for the deposition of the sensing-layer from a liquid dispersion of PbS-NCs in water directly on top of interdigitated electrodes (IDEs), Fig.\ref{Fig.Fabr}a-c. The sensitive layer was formed after post-processing the deposited PbS-NCs through a multi-step dry thermal process at mild-temperatures. The microstructure of the highly porous sensitive layer of PbS-NCs, obtained after post-processing, is shown in Fig. \ref{Fig.Fabr}b. The inset shows an atomically-resolved image of a single PbS-NC showing a high degree of crystallinity. The typical dimension of PbS-NCs is the range of 15-25 $nm$ after the heating step in vacuum at 180 $\celsius$ for 30 minutes. Two sensors, named \textsf{\textbf{sv}} and \textsf{\textbf{sa}}, were produced with differently tuned sensitivity by alternating thermal annealing steps in vacuum or in open-air, respectively. After drop casting the PbS-NCs on the IDEs, initially, both samples underwent a two-step vacuum-assisted annealing, by heating at 150 $\celsius$ for 40 minutes and subsequently the temperature was increased to 180 $\celsius$ and maintained for 30 minutes. In the sequence, sample \textsf{\textbf{sv}} was heat-treated in vacuum at 220 $\celsius$ for 30 minutes, while sample \textsf{\textbf{sa}} was heat-treated in open-air at 220 $\celsius$ for 30 minutes. The first annealing steps in vacuum (150-180 $\celsius$) are expected to reduce the inter-grain distance and improve crystallinity, while the second step (the heat treatment) is expected to promote specific surface modifications. More precisely, the vacuum-assisted annealing is expected to promote deep interface (donor-like) states related to excess Pb atoms at the NCs surfaces, while heating in ambient air is expected to promote the formation of a thin polar layer of oxygenated species (e.g. \ch{PbO}, \ch{PbO2} and \ch{PbSO4}) related to surface trap states \cite{Choi2014,Giansante2017}.

\subsubsection*{Sensing layer}\label{Composition}

\begin{figure}[h!]
\centering
{\includegraphics[width=1\columnwidth]{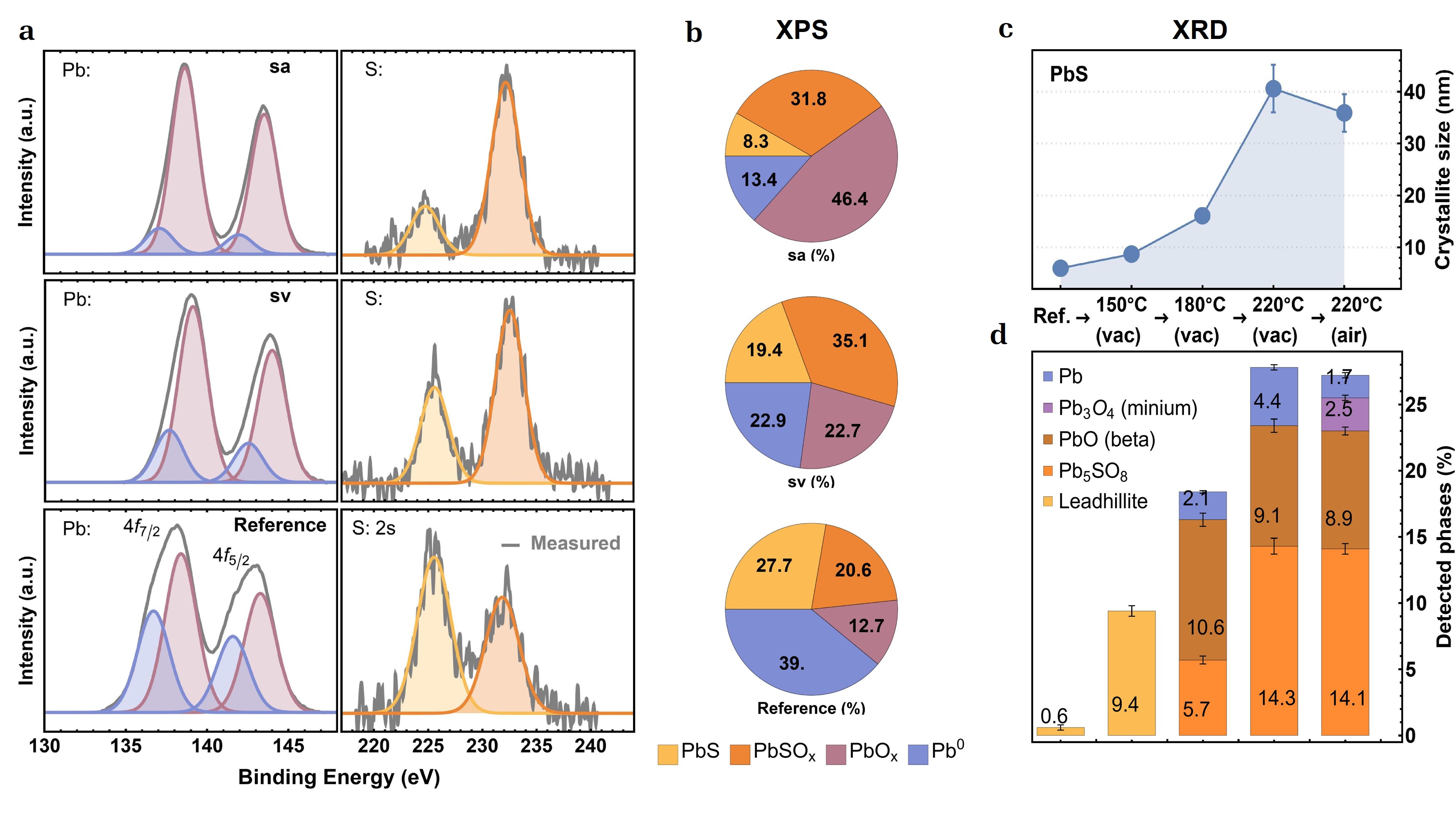}}
\caption{(a) XPS data used in the composition analysis, of lead-compounds (Pb: 4f$_{7/2}$ and 4f$_{5/2}$) on the left column, and for sulfur-compounds (S: 2s) on the right column (the gray curves are the raw experimental data). (b) Surface composition, in percentage of lead-compounds (\%) detected per sample. The amounts were quantified from the respective areas of the peaks shown in panel a. (c) Evolution of the crystallite size, on the PbS-NCs layer, after each thermal step, as determined from the XRD spectra (shown in the supporting information). (d) Composition (in \%) of the crystalline phases detected after each step indicated in (c), as determined from the Rietveld analysis of XRD spectra (the PbS phase was omitted to highlight the evolution of the composition). }\label{Fig.Comp}
\end{figure}

The structure and surface composition of the deposited PbS-NCs layers, corresponding to sensors \textbf{\textsf{sv}} and \textbf{\textsf{sa}}, was analyzed by X-ray photoelectron spectroscopy (XPS). In addition, a third sample, not subjected to any post-deposition treatment, was also analyzed and used as the reference. For the quantitative analysis of XPS data, the peaks were identified in the \textit{lead}-Pb(4f) and \textit{sulfur}-S(2s) regions of the narrow spectrum scans, as indicated in Fig.\ref{Fig.Comp}a. The peak positions corresponding to the identified chemical species for each sample are displayed in Table \ref{tab.XPS}. The surface amounts of lead-sulfide (\ch{PbS}), sulfates (\ch{PbSO_x}), oxides (\ch{PbO_x}), and neutral-lead (\ch{Pb^0}), were quantified and the results are presented in Fig.\ref{Fig.Comp}b, where each row corresponds to one of the samples, i.e., \textsf{reference}, \textbf{\textsf{sv}}, and \textbf{\textsf{sa}}, respectively (from bottom-to-top). From Fig.\ref{Fig.Comp}b, the thermal process results in the oxidation of the initial \ch{PbS} and \ch{Pb^{0}} species, to promote the formation of sulfates (\ch{PbSO_{x}}) and oxides (\ch{PbO_{x}}). As expected, heating in open-air (\textit{as for sample} \textsf{\textbf{sa}}) promoted the formation of the highest amount of oxygenated species, 78.2 \%, against 57.8 \% when heated in vacuum (\textit{as for sample} \textsf{\textbf{sv}}). Another fundamental difference to highlight between the two samples is the presence of sulfur-richer surface on the sample treated in vacuum (\textsf{\textbf{sv}}), i.e. 54.5 \% of sulfur species (\ch{PbS} and \ch{PbSO_{x}}) compared to 40.1 \% on the sample heated in open-air (\textsf{\textbf{sa}}).

\begin{table}[h!]
\caption{XPS peak attributions.} \label{tab.XPS}
\centering
\begin{tabular}{lccc}
\hline
Samples:  & Reference & \textbf{\textsf{sv}} (vac) 220 $\celsius$-30 min. & \textbf{\textsf{sa}} (open-air) 220 $\celsius$-30 min. \\
 peak positions: & (eV) & (eV) & (eV) \\
\hline
Pb$^{0}$:$(4f_{7/2})/(4f_{5/2})$ & 136.7/141.58  &  137.65/142.53  & 137.09/141.97  \\
PbO$_{\mathrm{x}}$:$(4f_{7/2})/(4f_{5/2})$ & 138.39/143.27  &  139.11/143.99  &  138.64/143.52 \\
PbS:$(2s)$ & 225.55  &  225.61  & 224.71 \\
PbSO$_{\mathrm{x}}$:$(2s)$ & 231.84  &  232.56  & 232.13 \\
\hline
\end{tabular}
\end{table}

The XRD spectra was analyzed using the Rietveld refinement technique to characterize the synthesized NCs, and quantify the composition of crystalline phases and the effect of the thermal process steps. The corresponding evolution of the mean crystallite size in the PbS-phase is presented in Fig.\ref{Fig.Comp}c. The analysis confirmed the role of the thermal treatment steps in promoting the crystalline quality of the PbS-NCs. Before any thermal treatment the mean crystallite size was determined as 6 $\pm$ 0.1 nm, while when heated under vacuum, between 150 $\celsius$ and subsequently to 180 $\celsius$, the NCs size increased systematically up to 16.1 $\pm$ 0.7 nm, while it increased sharply to 40.6 $\pm$ 4.6 nm, after annealing in vacuum at 220 $\celsius$. While the initial effects of annealing in vacuum is to promote a better crystallinity on the PbS-core, the sharp increase observed after annealing at 220 $\celsius$ suggests the promotion of partial agglomeration and fusion of PbS-NCs. Subsequently, after the last step, by heating in ambient-air at 220 $\celsius$, one observes the shrinkage of the crystallite size to 35.9 $\pm$ 3.6 nm, due to the oxidation of the PbS-NCs, with the formation of oxygenated species on the surface \cite{Sadovnikov2011,MOZAFARI2012}; confirmed by the XPS and XRD results. In Fig.\ref{Fig.Comp}d is presented the evolution of composition, with the crystalline phases quantified after each step (the PbS phase amount was omitted to highlight the evolution of the composition and can be inferred from the difference in \%). 

The chemical route driving the composition evolution of the sensing layer is proposed as follows: After deposition by drop coating from liquid phase in water, before any thermal treatment, the PbS-NCs layer is exposed to ambient-air at room-temperature; the surface begins to react with atmospheric oxygen and the oxidation of \ch{PbS} yields the formation of \ch{PbSO4}, while the adsorbed \ch{CO2} can react to form \ch{CO3^2-}, at high pH, yielding a carbonate phase (\ch{PbCO3}). The presence of $\mathrm{Pb(OH)_{2}}$ is expected, as the OH-groups strongly binds to the Pb-atoms at the surface and help to stabilize the PbS nanoparticles. First, heating at 150 $\celsius$ promotes crystallization and the formation of a leadhillite phase, \ch{Pb4SO4(CO3)2(OH)2}; that alternates lead-carbonate and lead-sulfate sheets (\ch{PbSO4}-\ch{Pb(OH)2}-2\ch{PbCO3}). The leadhillite phase is expected to thermally decompose forming anglesite (\ch{PbSO4}) and lanarkite (\ch{Pb2(SO4)O}). In turn, the lanarkite phase is likely to thermally decompose forming \ch{PbO} and releasing \ch{SO3}. This decomposition mechanism was confirmed by the thermal stability analysis, in Fig.S1 (in the supporting information), where the pair of peaks observed at 173 $\celsius$ (64, 48 amu) can be attributed to the fragmentation of \ch{SO3^+} (an unstable ion), via ionization in the mass spectrometer, forming \ch{SO2^+} and \ch{SO^+}. The formation of a metallic-Pb phase, as observed in vacuum, could be related to a reaction involving the residual carbon as a reducing agent, via the reaction \ch{PbO + C} $\rightarrow$ \ch{Pb + CO} (in standard conditions, this reaction is expected to become spontaneous at approximately $298 \celsius$). The detected \ch{Pb5SO8} phase stems from the formation of a solid mixture 4PbO.PbSO4, in vacuum, whereas the minium phase, \ch{Pb3O4}, is expected to form in open-air \cite{Milodowski1984}. The minium mineral corresponds to a mixed valence compound (2\ch{PbO.PbO2}) where the lead-atoms can present different oxidation states, \ch{Pb^2+} or \ch{Pb^4+}. The trend observed on the evolution of the \ch{Pb5SO8} phase amount, in Fig.\ref{Fig.Comp}d, follows closely the evolution of the crystallite size of the PbS phase (Fig.\ref{Fig.Comp}c). During the oxidation in open-air, at 220 $\celsius$ for 30-minutes, the oxidized species remain stable (i.e. $\beta$-PbO and \ch{Pb5SO8}) while the metallic-Pb at the surface is oxidized to yield the minium phase (\ch{Pb3O4}).

\subsection*{Sensing characteristics}

\begin{figure}[h!]
\centering
{\includegraphics[width=0.5\columnwidth]{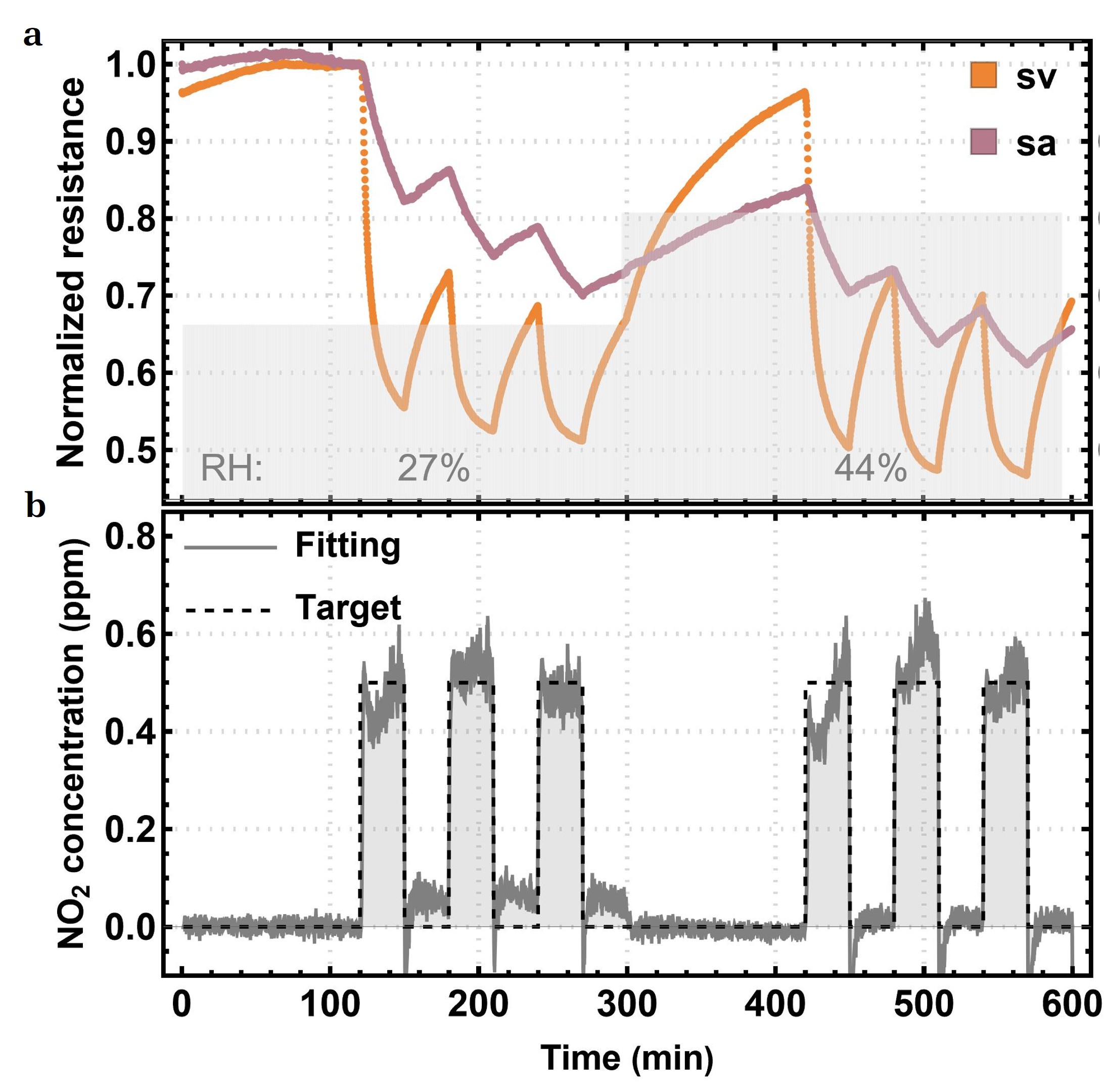}} 
\caption{Sensing characterization: (a) Variation of the normalized resistances of sensors \textsf{\textbf{sv}} and \textsf{\textbf{sa}}, at room-temperature (RT) and variable relative humidity (RH), during the measurement under a constant flow of synthetic air with introduction of step-like concentrations of \ch{NO2} gas  (\textit{the target curve in panel b, black dashed line}). (b) \textit{In gray} (shaded area), the \ch{NO2} concentration calculated from the resistances variations (\textit{in panel a}), obtained after fitting of equation \ref{eq:dconv}, using the parameters in Table \ref{tab.Calib}. }\label{Fig.Sens}
\end{figure}

The gas sensing characteristics of the sensors pair, \textbf{\textsf{sv}} and \textbf{\textsf{sa}}, were evaluated by exposing the sensors to predetermined (step-like) concentrations of \ch{NO2} gas, at regular intervals, in a constant flow of synthetic-air. During this test,  two measurement series containing \ch{NO2} gas were applied in concentration pulses of 0.5 ppm, in two different values of relative humidity (RH), 27 \% and 44 \%, respectively. The electrical resistances of both sensors was continuously monitored and recorded during the test, and the result is presented in Fig.\ref{Fig.Sens}-a. One can observe a higher sensitivity and faster reaction/recovery (i.e., improved sorption kinetics) on sensor \textbf{\textsf{sv}}, compared to sensor \textbf{\textsf{sa}}. In addition, the higher humidity environment promoted a faster recovery on sample \textbf{\textsf{sv}} while sample \textbf{\textsf{sa}} is not significantly affected by the humidity variation; as shown in the comparison of the estimated time constants of reaction/recovery, in Tab.\ref{tab.kinet}. The observed behavior, and especially the distinct effect of humidity in promoting \ch{NO2} desorption on sample \textbf{\textsf{sv}}, suggests the predominance of distinct reaction mechanisms in each sample; in accordance with the specific device models (described in a companion paper \cite{Fernandes2026}), where the conductivity of the sensing layer is assumed to be dominated by the built-in intergranular potential barriers (that can form at the interface of connected nanocrystals). The slower reaction observed on sample \textsf{\textbf{sa}}, in Fig.\ref{Fig.Sens}a, is the manifestation of a surface with lower sticking probability in comparison with sample \textsf{\textbf{sv}}. This can be related to longer surface diffusion lengths, as in a partially passivated surface. On the other hand, the slower recovery observed on sample \textsf{\textbf{sa}}, compared to sample \textsf{\textbf{sv}}, can be generally associated to a dominant presence of interaction sites with higher binding energy at the surface. These hypothesis will be further tested in the remainder of the paper, combining XPS, XRD analysis and DFT calculations.

\begin{table}[h!]
\centering
\captionsetup{width=.5\textwidth}
\caption{Time constants $(\tau)^{*}$ in minutes ($\pm 15\%$), for the response curves in Fig.\ref{Fig.Sens}-a} \label{tab.kinet}
\begin{tabular}{ |p{2cm}|p{2cm}|p{2cm}|  }
\hline
\textsf{\textbf{sv}} (reac./recov.) & \textsf{\textbf{sa}} (reac./recov.) & RH (\%) \\
\hline
 8/25  &  31/190 & 27  \\
 8/20  &  31/210 & 44   \\
\hline
\end{tabular} \\
\begin{tabular}{ p{7cm}  }
\textbf{*} estimated using the exponential decay fitting, B $\pm A(1-exp[-t/\tau])$.
\end{tabular} \\
\end{table}

\begin{subequations}
\begin{align}
\delta_{b}(r_{sv},r_{sa}) = \frac{r_{sv}}{r_{sa}^b} \label{eq:dconva}  \\
\Gamma(\delta_{b}, \dot \delta_{b}) =-\frac{D \ln \delta_{b}+\frac{\dot \delta_{b}}{\delta_{b}}}{A\left[a+\ln \delta_{b}\right]}\label{eq:dconvb}
\end{align}
\label{eq:dconv}
\end{subequations} 

An important aspect for the general applicability of gas sensors is the ability to quantitatively extract the gas-concentration in real-time, and with satisfactory accuracy. In the following, we apply a sensing scheme based on the relative resistance measurement, using the pair of devices, \textsf{\textbf{sv}} and \textsf{\textbf{sa}}. The method consists in combining raw chemiresistance signals from two sensors placed in the same environment and use the conversion formula (\ref{eq:dconv}), derived by authors of this paper as described in detail in ref. 32 (in preparation). It allows to extract the gas-concentration, $\Gamma (t)$, by using as inputs the relative differential-resistance variation in the sensor-pair, $\delta_{b}(r_{sv},r_{sa})$, and its time-derivative $\dot \delta_{b}(r_{sv},r_{sa})$, where $r_{sv}$ and $r_{sa}$ are the normalized resistances of each device in the pair (i.e., normalized by their respective baseline-resistances values, $R_{0\{sv\}}$ and $R_{0\{sa\}}$, that are measured when the device is exposed to a neutral atmosphere). The parameters, $A$ and $D$, are defined by the sorption kinetics; representing the adsorption and desorption coefficients, respectively. While the parameters $a$ and $b$ are closely related to each other and dependent on the device model \cite{Fernandes2026}. 

The conversion formula (\ref{eq:dconv}) was applied to the chemiresistance dataset, for samples \textsf{\textbf{sv}} and \textsf{\textbf{sa}}, in panel Fig.3-a, and the parameters ($a$, $b$, $A$, $D$) were obtained by fitting of $\Gamma (t)$ to the target (\ch{NO2}) concentration applied during the test. The result of the fitting and the comparison with the target gas concentration during the test is presented in Fig.\ref{Fig.Sens}-b. The fitting was performed independently, for the two regions with different values of RH, and the obtained parameters are presented in Table \ref{tab.Calib}. At higher RH, the values obtained for coefficients $A$ and $D$ are also higher; which aligns with the improved sorption kinetics observed in sample \textsf{\textbf{sv}}, and its enhancement at higher humidity. Overall, one can observe that the conversion formula can match the presence of a step-like concentration of \ch{NO2} gas, without delay, and in close quantitative agreement with the target gas profile during the test.

\begin{table}[h!]
\centering
\captionsetup{width=.5\textwidth}
\caption{Parameters obtained from fitting of eq.(\ref{eq:dconv}) on the target concentration (\textit{black-dotted curve}), Fig.\ref{Fig.Sens}b.$^*$} \label{tab.Calib}
\begin{tabular}{lcccccc}
\hline
\textit{\textbf{a}} & \textit{\textbf{b}} & \textit{A} & \textit{D} & $R_{s0}$ & $R_{r0}$ & RH \\
\hline
 0.60  &  0.53  &  0.336 & 0.040 & 6.6 $M\Omega$ & 3.0 $M\Omega$ & 27\% \\
 0.60  &  0.53  &  0.403 & 0.044 & 6.6 $M\Omega$ & 3.0 $M\Omega$ & 44\% \\
\hline
\end{tabular} \\
$^*$for room-temperature.
\end{table}

\subsubsection*{DFT}\label{DFT}

\begin{figure}[ht]
\centering
\includegraphics[width=0.5\columnwidth]{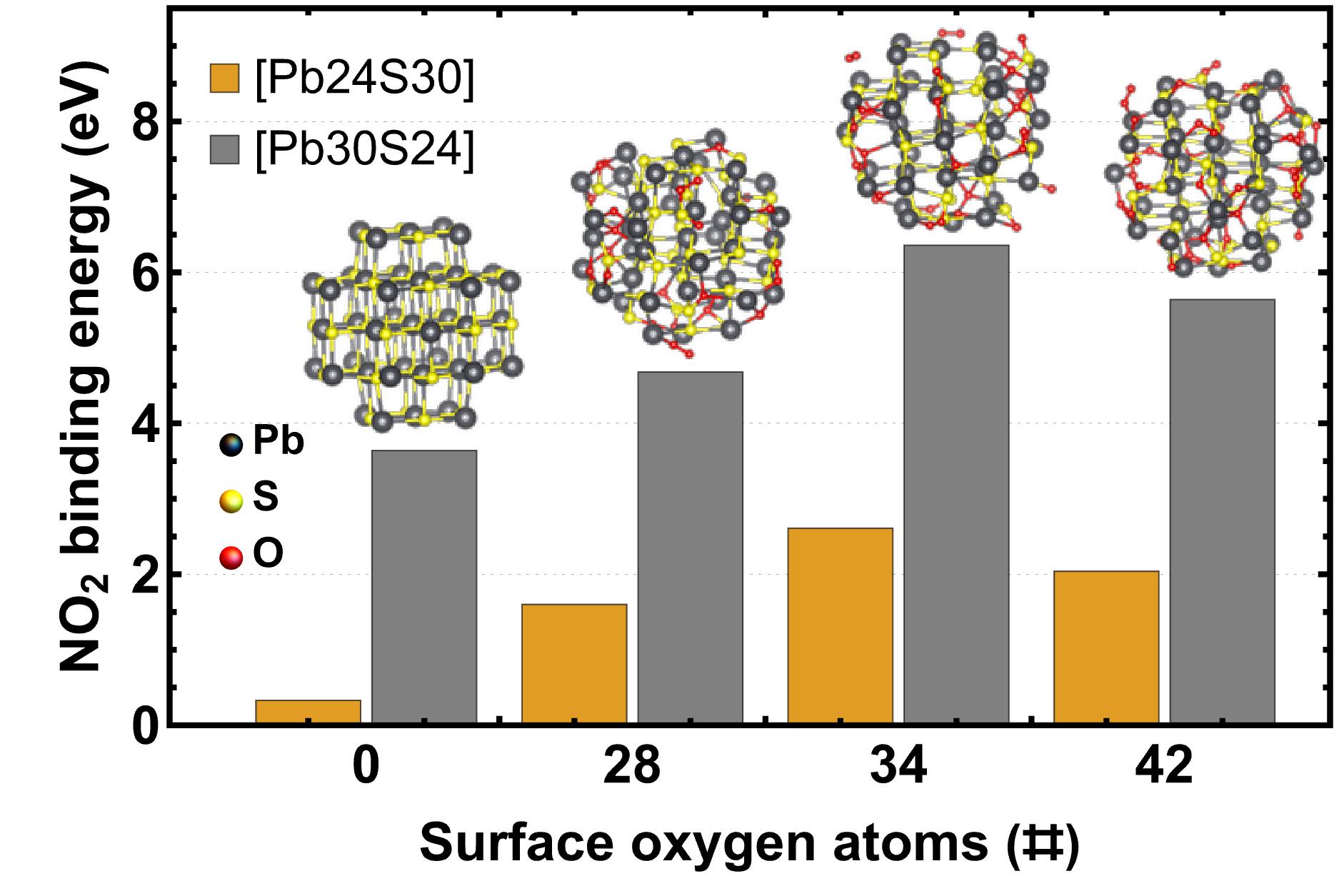}
\caption{DFT: Results of theoretical analysis, based on DFT calculations, exploring the binding energy of \ch{NO2} molecule on the surface of a model particle as a function of the surface composition, i.e. combining different amounts of lead-to-sulfur, and decoration with oxygen atoms (each value of binding energy corresponds to an averaged value taken from 6-different interaction sites, randomly distributed on the surface). This result allowed to clarify the role of the ratio Pb-to-S and the oxidation level, in balancing the surface reactivity and the sorption kinetics.}\label{Fig.DFT}
\end{figure}

In order to better understand the interplay between surface composition and the sensing characteristics of \ch{NO2} observed experimentally (Fig.\ref{Fig.Sens}a), we turned to Density Functional Theory (DFT) simulation of the interaction of \ch{NO2} molecules with PbS nanoparticles. The focus of the analyses was on the role of surface oxidation and the lead-to-sulfur stoichiometry, in promoting or preventing interaction with \ch{NO2}. The calculations were performed on PbS model-particles formed by a cluster of 81-atoms in total (of Pb and S), featuring 54-atoms on the external layer (with dangling bonds). These particles were decorated with an increasing number of oxygen atoms (0, 28, 34, and 42), as represented in Fig.\ref{Fig.DFT}. These Finite-size clusters are used to probe local surface chemical interactions, not size-dependent or edge-dominated effects.  To study the effect of the lead-to-sulfur surface stoichiometry, the clusters were created in pairs with swapped lead-sulfur atomic positions, yielding different surface terminations; i.e., lead-rich (surf-[\ch{Pb30S24}]) or sulfur-rich (surf-[\ch{Pb24S30}]), corresponding to the gray and orange color in Fig.\ref{Fig.DFT}). The mean binding energy for an interacting \ch{NO2} molecule was obtained dependent on the number of surface oxygen atoms and surface terminations of the clusters, and the values are presented in Fig.\ref{Fig.DFT}. Each value of binding energy in Fig.\ref{Fig.DFT} corresponds to the mean value of independently calculated \ch{NO2} molecule interacting with 6 different interaction-sites, randomly distributed on the cluster surface. These results point towards the role of sulfur compounds in reducing the binding energy, as all of the clusters with S-rich surfaces ([\ch{Pb24S30}]) show lower binding energy towards \ch{NO2}. This result is consistent with the faster desorption (recovery) that was experimentally observed for sensor \textsf{\textbf{sv}} (Fig.\ref{Fig.Sens}a.), and correlates to the presence of a sulfur-richer surface in this sensor (as determined by XPS, in Fig.\ref{Fig.Comp}b). Regarding the surface oxidation degree, the results show an interesting behavior that is also consistent with the experimental sorption dynamics; the \ch{NO2} mean binding energy consistently increased with the number of oxygen atoms decorating the surface, reaching the maximum for 34 oxygen atoms. However, it decreased when surface oxygen content (oxidation) increased from 34 to 42 atoms; which can be related to the presence of lead in the \ch{Pb^{4+}} oxidation state. This trend is consistent with the passivation effect observed in sensor \textsf{\textbf{sa}} (Fig.\ref{Fig.Sens}a), and can be associated to the formation of minium phase (Fig.\ref{Fig.Comp}d). Overall, the DFT simulations reflect qualitative trends in \ch{NO2}-surface interactions through idealized PbS clusters and effective surface oxidation, omitting explicit modeling of grain boundaries, hydroxylation, or complete oxide/sulfate phases. The calculated binding energies should be taken as relative indications of \ch{NO2} affinity, not absolute adsorption energies.

\subsubsection*{Sensing mechanisms}\label{sens.mec}

The general sensing mechanism in chemiresistive sensors involves the formation of a space-charge region (electron-depletion/hole-accumulation layer in n/p-type semiconductor) due to adsorbed oxygen, and hydroxylated lead, when the sensing layer is exposed respectively to air and humidity (as detailed in the introduction). \cite{Wang2010,Ji2019,Nikolic2020} The correlation between the surface composition (Fig.\ref{Fig.Comp}b-d) in comparison to the observed respective sorption dynamics on both samples (Fig.\ref{Fig.Sens}a), defined by the time constants of reaction/recovery in Tab.\ref{tab.kinet}, seem to point towards the role of the $\beta$-\ch{PbO} phase in promoting stronger interaction with \ch{NO2}, making it harder to desorb, and the predominance of $metalic$-\ch{Pb} when the surface reaction yelds faster sorption kinetics, along with the role of humidity in promoting desorption. Furthermore, the surface amount of \ch{Pb3O4} phase, with high oxidation order (\ch{Pb^{4+}}), can be viewed as passivated (\ch{Pb^{0}}) interaction sites, i.e., reducing the \ch{NO2} sticking probability and yielding the longer reaction times observed for sample \textbf{\textsf{sa}} (Tab.\ref{tab.kinet}). Overall, the DFT results seem to confirm the role of \ch{PbO} and \textit{metallic}-\ch{Pb} in promoting the surface reaction with \ch{NO2}, i.e. the role of lead in the oxidation states \ch{Pb^{2+}} and \ch{Pb^{0}}, respectively, as interaction sites, and the surface passivation when saturated with oxygen that can be associated with the formation of \ch{Pb^{4+}}. Furthermore, the presence of sulfur compounds (\ch{PbS} and \ch{PbSO_{x}}) on the surface (Fig.\ref{Fig.Comp}b) is expected to result in weakening of the interaction with \ch{NO2}, as confirmed by the DFT analysis comparing the binding energies of sulfur-rich and lead-rich surfaces. Therefore, the sorption dynamics (response/recovery times) is defined by the surface composition, in particular regarding the density and the surface ratio of lead in the oxidation states \ch{Pb^{2+}} and \ch{Pb^{0}}. The proposed sensing mechanism is tentatively associated to the formation of adsorbed lead (unidentate) nitrate via two different (surface-mediated) reactions, \ref{eq:PbO} and \ref{eq:Pb}; based on the direct interaction between \ch{NO2} and surface \ch{PbO} or intermediate by a lead-hydroxyl (\ch{Pb-OH}) layer (formed by hydroxylation of metallic-lead) and adsorbed oxygen:\cite{Baltrusaitis2012,Shi2021}

\begin{subequations}
\begin{align}
\ch{_{(surf)}PbO} + \ch{NO2_{(g)}} \longrightarrow \ch{PbNO3_{(ads)}} \label{eq:PbO}  \\
2\ch{_{(surf)}Pb(OH)} + 2\ch{NO2_{(g)}} + 1/2\ch{O2_{(ads)}} \longrightarrow 2\ch{PbNO3_{(ads)}} + \ch{H2O_{(ads)}} \label{eq:Pb}
\end{align}
\label{eq:convform}
\end{subequations}

Thus, one can assume the reaction \ref{eq:PbO} as the predominant reaction on sample \textbf{\textsf{sa}}, i.e. yielding the slower recovery due to higher binding energy, resulting from the higher amount of \ch{PbO}. In contrast, the slower reaction on sample \textbf{\textsf{sa}} is attributed to passivated sites, as a consequence of formation of \ch{Pb3O4} phase. The reaction \ref{eq:Pb} is predominant for the faster reaction, and the effect of enhanced recovery observed at higher humidity, on sample \textbf{\textsf{sv}}, due to to the predominance of metallic-\ch{Pb} over \ch{PbO} on the surface. Therefore, on both samples, one can assume the presence of, at least, two-types of interaction sites, corresponding to the surface amounts of \ch{PbO} and \ch{Pb^{0}} that are also consistent with the detailed device models, for sensors \textbf{\textsf{sa}} and \textbf{\textsf{sv}}, presented in the companion paper \cite{Fernandes2026}.

\subsection*{CONCLUSIONS}\label{Conclusion}

We demonstrate a simple multistep dry-thermal process that can be designed in order to engineer the sensitivity and sorption kinetics of \ch{NO2} gas sensing, based on a layer of oxidized PbS-NCs; i.e. depending on the surface composition ratio of different oxidation states of lead (\ch{Pb^0}, \ch{Pb^{2+}}, \ch{Pb^{4+}}). A detailed sorption mechanism was described, and a close correlation between the surface composition and experimental sensing characteristics was established. In addition, the transduction mechanism (resistance control mechanism) was validated, based on the modulation of the intergrain potential barrier formed at the interface of connected PbS-NCs; by the application of a sensing scheme and a conversion formula, suitable for extracting the gas concentration (quantitatively and fairly accurately) from the raw resistance variation during gas sensing test.

The sorption mechanism is based on the formation of adsorbed lead (unidentate) nitrate on the surface of an oxidized layer of PbS-NCs. The sensing characteristics are defined by the lead-to-sulfur surface stoichiometry and the presence of surface ratio amounts of \textit{metallic}-\ch{Pb}, \ch{PbO}, and \ch{PbO2}. In addition, DFT calculations were performed to clarify the effect of distinct lead-to-sulfur surface stoichiometry, and different oxidation levels, on the surface reactivity to \ch{NO2}. The proposed sensing mechanism can correlate the surface composition of each sample, as determined in the XPS and XRD analysis, and the results obtained with DFT, in a consistent framework to explain the sensitivity and sorption dynamics observed experimentally on both samples. However, a complete physical description of the mechanisms of sensing is still incomplete, i.e., future work should aim at reaching a physical model that quantitatively correlates the surface reactions to the corresponding variation of the intergrain potential barrier.

In addition, here we proposed a simple and cost-effective route for the mass production of \ch{NO2} microsensors, that can operate at room-temperature and are able to quantitatively estimate low \ch{NO2} gas concentrations (below 1 ppm), i.e., for relevant applications in environmental monitoring and indoor air-quality control. The simple fabrication protocol is based on the deposition of PbS-NCs, directly from a liquid dispersion in water, followed by the activation of the sensitive layer using a combination of dry-thermal processes at mild temperatures, i.e., eliminating the need of chemicals that can either be process-incompatible and harsh to the environment. \cite{Balazs2018,Tang2019,Hu2023}.

\subsection*{METHODS}\label{methods}

\textbf{PbS nanoparticles:} The PbS-NCs were obtained via a simple wet chemical route in water, at room-temperature, using \ch{Pb(NO3)2} (Merk product number: 203580-10G) and \ch{Na2S} (Merk product number: 407410-10G) as precursors and 2-Mercaptoethanol (Merk product number: M6250-100ML) as the surfactant \cite{Mosahebfard2016,Nabiyouni2012}. Initially, 50 mL solution of \ch{Pb(NO3)2} with concentration 0.1M is poured into a triple-neck round bottom flask, and a volume of 100 mL of 2-Mercaptoethanol solution, with concentration 0.1M, is then added dropwise, followed by 50 mL of \ch{Na2S} solution, with concentration of 0.1M, that was slowly added. The mixture was vigorously stirred until the color of the final suspension turned black. The process was done at room-temperature using Milli-Q water in the solutions (that was previously degassed in argon for 15 minutes). A constant pressure of argon was permanently maintained, in the flask, in all stages during the synthesis. As a result, this synthesis protocol produces a dispersion of PbS-NCs with concentration of approximately 5 mg/mL. \\ \\
\textbf{Interdigitated electrodes:} Commercial IDEs-array chips (CMOSEnvi\textsuperscript{\textregistered}) were used as substrates for samples production, supplied by the company VOCSens \cite{VOCSens}. A total of $8$x IDEs on each chip were coated with PbS-NCs to form individual sensors. The IDEs are formed by thin-film strips consisting of Ti:Au layers (200:2000 \text{\AA} of thickness), with 2 $\mathrm{\mu m}$ wide and spaced by a 5 $\mathrm{\mu m}$ gap. The metal-strips are deposited on top of a $\mathrm{SiO_2}$ insulating layer. The area of each micro-electrode pair ranged from 50 $\mathrm{\mu m}$ x 350 $\mathrm{\mu m}$, on the smallest, to 500 $\mathrm{\mu m}$ x 1050 $\mathrm{\mu m}$, on the biggest pair (Fig.\ref{Fig.Fabr}c). All the samples used in this work were fabricated using the IDE with an area of  $50\mu m$ x $350\mu m$, corresponding to the smallest IDE (as detailed in the inset of Fig.\ref{Fig.Fabr}c). \\ \\
\begingroup
\textbf{Deposition of PbS-NCs on the IDEs:} For the sensors fabrication, the deposition of PbS-NCs was based on the drop drying method. For the deposition, the liquid-dispersion (Ink) of PbS-NCs in water (5mg/mL, as obtained after the synthesis) was subjected to an ultrasonic bath at mild-power (70-80W) before being filtered in a sequence using syringe-filters with 5.0 µm, 2.7 µm and 1.2 µm pores. The filtered ink was then loaded in a micro-fine syringe, with a 0.3 mL reservoir. The syringe needles must be cut vertically to form a perfect perpendicular cut at the extremity of the needle. The diameter of the needle is approximately ~400 µm. With this technique a droplet with sub-nanoliter volume can be formed (manually) in a semi-controlled way. The syringe is mounted on an old-wire bonder station to provide stability and mechanical precision on the 3-axis. In addition, the Wire bonder is equipped with a stereo microscope for sample visualization. The sample is placed on a sample holder, that is equipped with a ceramic heater and thermocouple. During the drop casting the sample is heated at 65-80 \textcelsius \ to increase the water-evaporation rate and promote the formation of a homogeneous coating layer. After the drop casting, the samples are vacuum-dried at 70 \textcelsius \ for 2 h. \\ \\
\endgroup
\begingroup
\textbf{XPS:} The XPS analysis was performed on deposited layers of PbS-NCs on Au-coated \ch{SiO2} substrates. The XPS measurements were performed with a PHI 5000 VersaProbe III Photoelectron Spectrometer (Physical Electronics (USA)), equipped with a monochromatized micro focused Al K$\alpha$ X-ray source, powered at 50 W. The pressure in the analysis chamber was kept around 10-6 Pa. The angle between the surface normal and the axis of the analyzer lens was 45°. High resolution scans of the Pb4f and S2s photoelectron peaks were recorded from a spot diameter of 200 $\mu$m using pass energy of 13 eV and step size of 0.1 eV. Charge stabilization was achieved thanks to a combination of Argon and electron guns. Data treatment was achieved using the CasaXPS software (Casa Software Ltd, UK). The Shirley algorithm was applied to remove the background and the shape of the spectral peaks were modeled (fitted) using Gaussian-Lorentzian product formula (with 85\% Gaussian and 15\% Lorentzian). \\ \\
\endgroup
\begingroup
\textbf{XRD:} The ink of PbS-NCs was deposited by the drop-drying method on zero-background Si Holders. The XRD spectrometer (Bruker D8 Discover) was operated at 30 mA, using Cu K$\alpha$ (Ni-filtered) radiation source with a wavelength of 1.5406 Å, and equipped with a LynxEye detector. The goniometer was equipped with a fixed divergent slit opening of 0.6 mm and Soller slit with opening angle of 2.498° on the primary-beam, and anti-scatter slit opening of 6.76 mm and Soller slit opening of 4° on the secondary-beam. Data was acquired between 35° and 85° and collected with a step interval equal to 0.00831° over 2$\theta$, with a scan speed of 0.1 sec/step. The evolution of the crystallinity was determined after each step of thermal treatment by analyzing the XRD spectra, applying the Rietveld refinement technique using Profex software, to characterize and quantify the composition of crystalline phases (Fig.\ref{Fig.Comp}c-d). A total of 4-steps of thermal processes were applied sequentially to the XRD samples: \textbf{(1)} The first step of the thermal process consisted in a vacuum-assisted annealing at 150 $\celsius$ for 40 minutes. \textbf{(2)} The second step consisted of increasing the temperature to 180 $\celsius$, in vacuum, and waiting for 30 minutes before the next step. \textbf{(3)} In the third step, the temperature was increased again, in vacuum, up to 220 $\celsius$ and maintained for 30 minutes. Finally, in the last step \textbf{(4)}, the sample was removed from the vacuum-assisted annealing system and placed on a hot-plate, in open-air, and heated again at 220 $\celsius$ for 30 minutes. \\ \\
\endgroup
\begingroup
\textbf{Gas sensing:} For the gas sensing characterization, the samples were mounted inside small chambers, that are connected in series to a gas-flow line, and electrically connected to a digital multimeter (equipped with a multiplexed channel input). Small concentrations of the target gas ($\mathrm{NO_{2}}$) are mixed in the synthetic-air flow, taking the form of step-like pulses with duration of 30-minutes each and separated by a 30-minutes recovery period (Fig. \ref{Fig.Sens}b - \textit{black dotted curve}). Calibrated mass-flow controllers are used to maintain the constant flow of 2-liter/minute in the gas-line and a relative humidity (RH) in the gas mixture during measurement. The set-up (gas characterization bench) is composed by 4x independent gas-lines: (1) $\mathrm{O_{2}}$, (2) $\mathrm{NO_{2}}/\mathrm{N_{2}}$ (2.6 ppm), (3) \textquotesingle dry\textquotesingle-$\mathrm{N_{2}}$ and (4) \textquotesingle wet\textquotesingle-$\mathrm{N_{2}}$. The relative humidity is controlled using a bubbler (filled with DI-water) connected to an independent $\mathrm{N_{2}}$ gas line (\textquotesingle wet\textquotesingle-$\mathrm{N_{2}}$ line). A constant excitation current of $0.5$ $\mu A$ was applied on the samples electrodes during measurements, and the power dissipated per device was typically in the range $1$-$2.5 \mu W$. These operational characteristics corresponds to ultra-low power applications, where a small/compact energy source can be employed (e.g. in smart sensors applications).  \\ \\ 
\endgroup
\begingroup
\textbf{DFT:} The DFT calculations were performed using the projector augmented waves (PAW) method as implemented in the Quantum ESPRESSO (QE) package.\cite{Giannozzi2009} The exchange-correlation effects were treated using the Perdew-Burke-Ernzerhof (PBEsol) functional, which is optimized for solids and surfaces. Spin–orbit coupling effects are also included in our calculations. The QE library provided the \text{Pb.pbesol-dn-rrkjus\_psl.1.0.0.UPF}, \text{S.pbesol-nl-rrkjus\_psl.1.0.0.UPF}, \text{N.pbesol-n-rrkjus\_psl.1.0.0.UPF}, and \text{O.pbesol-n-kjpaw\_psl.0.1.UPF} pseudopotentials. A plane-wave kinetic energy cutoff of 45 Ry and a charge density cutoff of 400 Ry were employed to ensure accurate results while maintaining computational efficiency. These settings were chosen based on convergence tests to balance precision and computational cost. The structural optimization was truncated after the Hellmann Feynman forces were less than 10-5 eV/Å. The cluster of 81 atoms has been cleaved from the rock salt (NaCl) crystal of the optimal lattice constant a = 6.003 Å of PbS as shown in Fig.\ref{Fig.DFT}. The crystal structure of the bulk was fully relaxed before and after slicing out the nanoparticles.  \\
\endgroup

\section*{Acknowledgement}



The authors discloses support for the research of this work from SPW Research [Convention n° 2010243, BEWARE 2020 - Appel 2] and the company VOCSens [n° BCE 0721.610.714]. B.H. (senior research associate) acknowledge support from the F.R.S.-FNRS. The authors also acknowledge financial support from the ARC project DREAMS (21/26.116).



\section*{Supporting information}


The following files are available.
\begin{itemize}
  \item Elucidating different \ch{NO2} sensing mechanisms in oxidized PbS nanocrystals: The multistep dry-thermal process was characterized and the results are made available in the supporting information. It contains the results regarding the thermal stability analysis of the deposited layer of PbS nanocrystals, and the detailed XRD spectra concerning the evolution of the composition and crystallinity of the PbS nanocrystals layer, after different thermal treatment steps.
\end{itemize}



\bibliography{Bibliography.bib} 

\newpage

\end{document}


\maketitle

\subsection*{Thermal stability analysis}\label{ThermStab}

The evolution of the deposited layer of PbS-NCs was studied through thermal stability analysis and X-ray diffraction (XRD), to identify the composition and evolution of crystalline phases after each step.
For the thermal stability analysis, a layer of PbS-NCs was heated in vacuum, up to 300 $\celsius$, at a constant rate and the mass spectra of desorbed chemical species was recorded over time. The more prominent desorption peaks were identified (after removing the background) and are presented in Fig.\ref{Fig-TS}. One can identify the first group of peaks appearing at about 157 $\celsius$ (60, 59, 45 amu), followed by a pair of peaks appearing at 173 $\celsius$ (64, 48 amu) and the last peak appearing close to 216 $\celsius$ (30 amu). The first group is attributed to the thermal degradation of the capping agent (2-Mercaptoethanol), used in the synthesis of the PbS-NCs, which can fragment respectively in $\mathrm{C_2H_4S, \ C_2H_3S, \ C_2H_5O}$ \cite{Nist-2Mercapt2025}. While the pair of peaks at about 173 $\celsius$ can be predominantly attributed to surface decomposition and desorption of $\mathrm{SO_{2}}$ and $\mathrm{SO}$. The last peak at 216 $\celsius$ can be attributed to a thermal degradation of the residual precursor $\mathrm{Pb(NO_{3})_{2}}$ and desorption of $\mathrm{NO_{2}}$ followed by its fragmentation yielding $\mathrm{NO}$ (i.e. caused by ionization to form $\mathrm{NO_{2}^{+}}$ followed by the loss of an oxygen to yeld $\mathrm{NO^{+}}$ and a neutral oxygen radical) \cite{Sayi2002}. \\

\begin{figure}[h!]
\setcounter{figure}{0}
\renewcommand{\figurename}{Fig.}
\renewcommand{\thefigure}{S\arabic{figure}}
\centering
{\includegraphics[width=0.5\columnwidth]{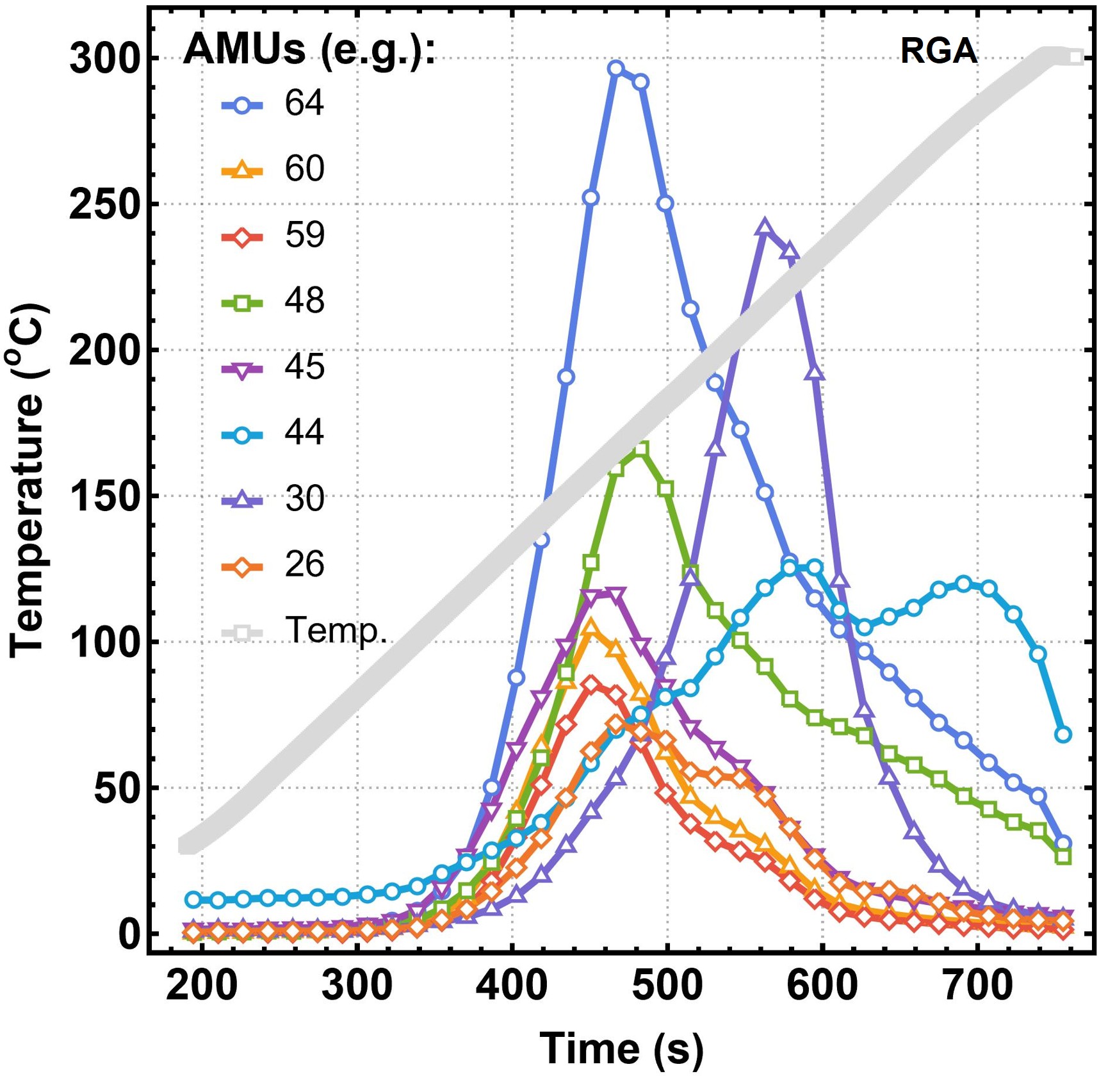}}
\caption{Thermal stability analysis: Atomic mass of desorbed species while heating the deposited layer of PbS-NPs up to 300 $\celsius$, at constant rate.}\label{Fig-TS}
\end{figure} 

\subsection*{Multistep dry-thermal process}\label{Multistep}

\begin{figure}[h!]
\renewcommand{\thefigure}{S\arabic{figure}}
\centering
{\includegraphics[width=0.5\columnwidth]{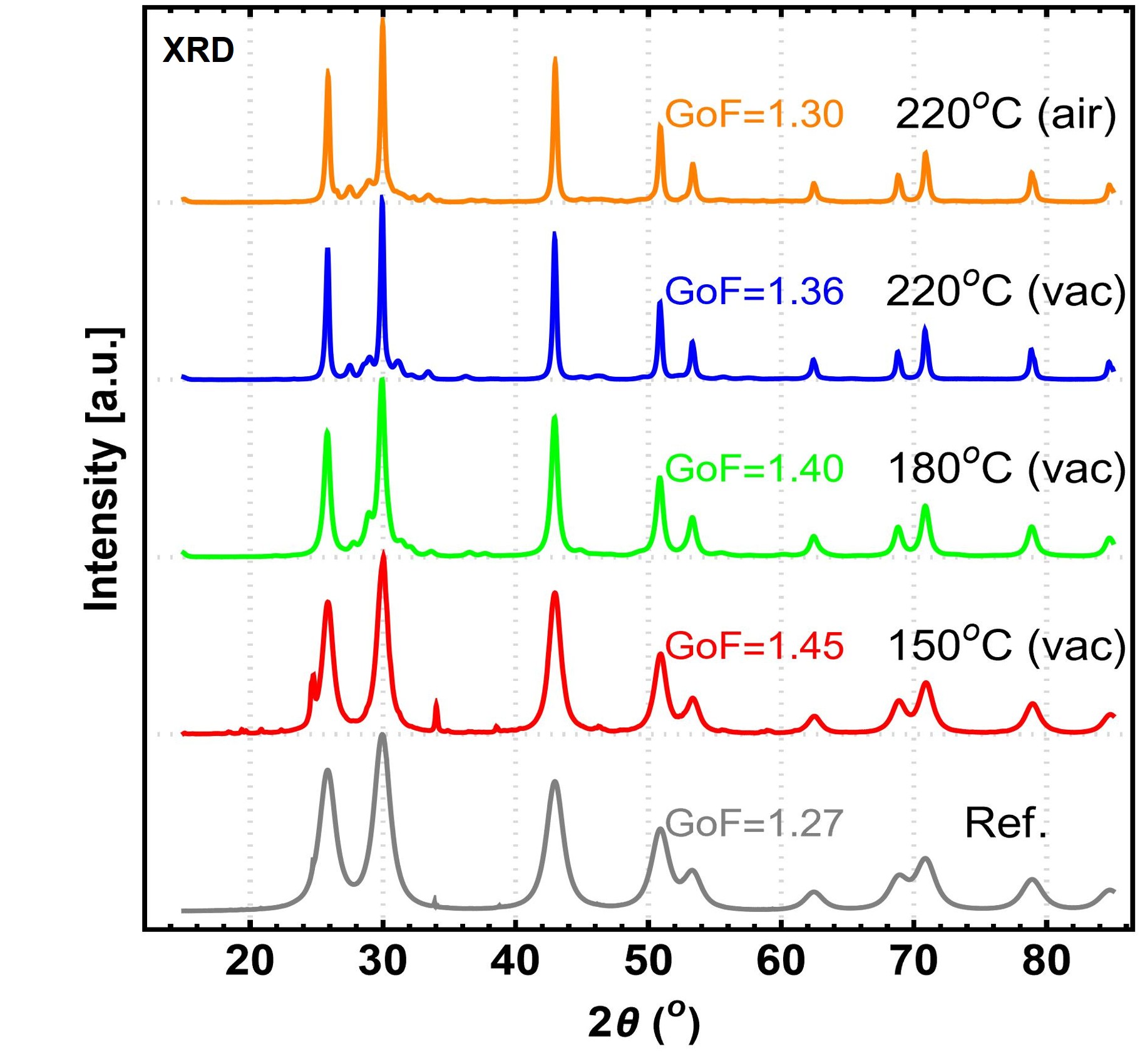}}
\caption{XRD: Each XRD spectrum measured after each thermal treatment step. The temperature is indicated, as well as the distinctive \textit{goodness-of-fit} (GoF) attributed to each result obtained from the Rietveld analysis}\label{Fig-XRD}
\end{figure}

The multistep thermal process was designed as a post-deposition activation protocol by applying dry thermal processes at mild temperature (i.e. operated below 300 $\celsius$) in line with standard CMOS back-end manufacturing technology. The evolution of the crystallinity was determined after each step of thermal treatment by analyzing the XRD spectra, in Fig.\ref{Fig-XRD}, using the Rietveld refinement technique to characterize and quantify the composition of crystalline phases (shown in the main text). A total of 4-steps of thermal processes were applied sequentially to the XRD samples: \textbf{(1)} The first step of the thermal process consisted in a vacuum-assisted annealing at 150 $\celsius$ for 40 minutes. \textbf{(2)} The second step consisted of increasing the temperature to 180 $\celsius$, in vacuum, and waiting for 30 minutes before the next step. \textbf{(3)} In the third step, the temperature was increased again, in vacuum, up to 220 $\celsius$ and maintained for 30 minutes. Finally, in the last step \textbf{(4)}, the sample was removed from the vacuum-assisted annealing system and placed on a hot-plate, in open-air, and heated again at 220 $\celsius$ for 30 minutes. \\


\newpage
\bibliography{SuppBibliography.bib} 







